# CONGESTION CONTROL IN COMPUTER NETWORKS: ISSUES AND TRENDS

Raj Jain
Digital Equipment Corp.
550 King St. (LKG 1-2/A19)
Littleton, MA 01460
Internet: Jain@Erlang.enet.DEC.Com

**Abstract**

Popular myths that cheaper memory, high-speed links, and high-speed processors will solve the problem of congestion in computer networks are shown to be false. A simple definition for congestion based on supply and demand of resources is proposed and is then used to classify various congestion schemes. The issues that make the congestion problem a difficult one are discussed, and the architectural decisions that affect the design of a congestion scheme are presented. It is argued that long-, medium-, and short-term congestion problems require different solutions. Some of the recent schemes are briefly surveyed, and areas for further research are suggested.

## 1 Introduction

Congestion control is concerned with allocating the resources in a network such that the network can operate at an acceptable performance level when the demand exceeds or is near the capacity of the network resources. These resources include bandwidths of links, buffer space (memory), and processing capacity at intermediate nodes. Although resource allocation is necessary even at low load, the problem becomes more important as the load increases because the issues of fairness and low overhead become increasingly important. Without proper congestion control mechanisms, the throughput (or *net*work) may be reduced considerably under heavy load.

In this paper, we begin with several myths about congestion and explain why the trend toward cheaper memory, higher-speed links, and higher-speed processors has intensified the need to solve the congestion problem. We then describe a number of proposed solutions and present a classification of congestion problems as well as their solutions. In Section 4 we explain why the problem is so difficult. In Section 5, we discuss the protocol design decisions that affect the design of a congestion control scheme. Finally, we describe our recent proposals and suggest areas for future research.

## 2 Myths About Congestion Control

Congestion occurs when the demand is greater than the available resources. Therefore, it is believed that as the resources become less expensive, the problem of congestion will be solved automatically. This has led to the following myths:

1. Congestion is caused by a shortage of buffer space and will be solved when memory becomes cheap enough to allow infinitely large memories.

2. Congestion is caused by slow links. The problem will be solved when high-speed links become available.

3. Congestion is caused by slow processors. The problem will be solved when the speed of the processors is improved.

4. If not one, then all of the above developments will cause the congestion problem to go away.

Contrary to these beliefs, without proper protocol redesign, the above developments may lead to more congestion and, thus reduce performance. The following discussion explains why.

*The congestion problem can not be solved with a large buffer space.* Cheaper memory has not helped the congestion problem It has been found that networks with infinite-memory switches are as susceptible to congestion as networks with low-memory switches [26]. For the latter, it is obvious that too much traffic will lead to buffer overflow and packet loss, as shown in Figure 1a. On the other hand, with infinite-memory switches, as shown in Figure 1b, the queues and the delays can get so long that by the time the packets come out of the switch, most of them have already timed out and have been retransmitted by higher layers. In fact, too much memory is more harmful than too little memory since the packets (or their retransmissions) have to be dropped after they have consumed precious network resources.

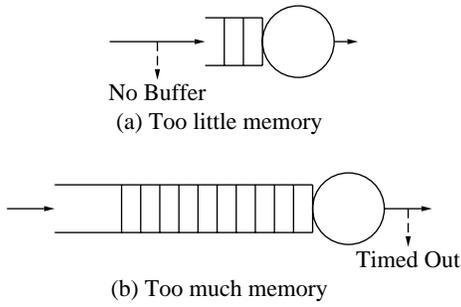

Figure 1: Too much memory in the intermediate nodes is as harmful as too little memory.

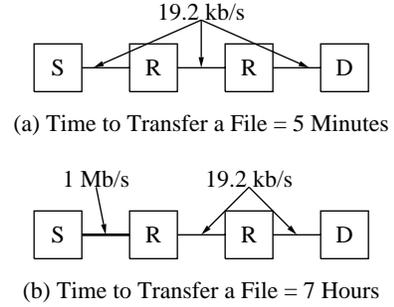

Figure 2: Introducing a high-speed link may reduce the performance.

*The congestion problem can not be solved with high-speed links*. In the beginning, the telephone links connecting computers had a speed of a mere 300 bits per second. Slowly, the technology improved, and it was possible to get dedicated links of up to 1.5 Mbits per second. Then came the local area networks (LANs), such as Ethernet with a speed of 10 Mbits per second. It was precisely at this point that the interest in congestion control techniques increased. This is because the high-speed LANs were now connected via a low-speed, long-haul links, and congestion at the point of interconnection became a problem.

The following experiment, although a contrived one, shows that introducing high-speed links without proper congestion control can lead to reduced performance [17]. Figure 2 shows four nodes serially connected by three 19.2 kbits per second links. The time to transfer a particular file was five minutes. After the link between the first two nodes was replace by a fast 1 Mbits per second link, the transfer time increased to seven hours! With the high-speed link, the arrival rate to the first router became much higher than the departure rate, leading to long queues, buffer overflows, and packet losses that caused the transfer time to increase.

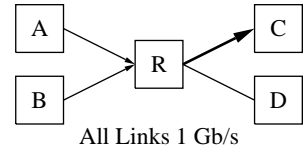

All Links 1 Gb/s

Figure 3: A balanced configuration with all processors and links at the same speed is also susceptible to congestion.

The point is that high-speed links cannot stay in isolation. The low-speed links do not go away as the high-speed links are added to a network. Introduction of high-speed links has increased the range of speeds that have to be managed. The protocols have to be designed specifically to ensure that this increasing range of link speeds does not degrade the performance.

*The congestion problem can not be solved with high-speed processors*. The argument for processors is similar to that for links. Introduction of a high-speed processor in an existing network may increase the mismatch of speeds and the chances of congestion.

*Congestion occurs even if all links and processors are of the same speed*. Our arguments above may lead some to believe that a balanced configuration with all processors and links at the same speed will probably not be susceptible to congestion. This is not true. Consider for example, the balanced configuration shown in Figure 3, where all processors and links have a throughput capacity of 1 Gbits per second. A simultaneous transfer of data from nodes A and B to node C can lead to a total input rate of 2 Gbits per second at the router R while the output rate is only 1 Gbits per second, thereby causing congestion.

The conclusion is that congestion is a dynamic problem. It cannot be solved with static solutions alone. We need protocol designs that protect networks in the event of congestion. The explosion of high-speed networks has led to more unbalanced networks that are causing congestion. In particular, packet loss due to buffer shortage is a *symptom* not a cause of congestion.

## 3  A Classification of Congestion Problems and Solutions

In simple terms, if, for any time interval, the total sum of demands on a resource is more than its available capacity, the resource is said to be congested for that interval. Mathematically speaking:

$$\Sigma \text{Demand} > \text{Available Resources} \qquad (1)$$

In computer networks, there are a large number of resources, such as buffers, link bandwidths, processor times, servers, and so forth. If, for a short interval, the buffer space available at the destination is less than that required for the arriving traffic, packet loss occurs. Similarly, if the total traffic wanting to enter a link is more than its bandwidth, the link is said to be congested.

The above definition of congestion, although simplistic, is helpful in classifying congestion problems as well as solutions. Depending upon the number of resources involved, a congestion problem can be classified as a single resource problem or a distributed resource problem, as shown in Figure 4. The single resource involved may be a dumb resource, such as a LAN medium, in which case, all the intelligence required to solve the congestion problem has to be provided by the users. Various LAN access methods, such as CSMA/CD (Carrier Sense Multiple Access with Collision Detection), token access, register insertion, and so on, are examples of solutions to the problem of single, dumb resource congestion. If the resource is intelligent, for example, a name server, it can allocate itself appropriately. The problem is more difficult if the resource is distributed as in the case of a store and forward network. For example, consider int the links as the resources, the user demands have to be limited so that the total demand at each link is less than its capacity. It is this set of problems dealing with distributed resource congestion that we are concerned with in this paper.

The simple definition of congestion above also allows us to classify all congestion schemes into two classes: those that dynamically increase the available resource, and those that dynamically decrease the demand. Some examples of both these types of schemes are described below.

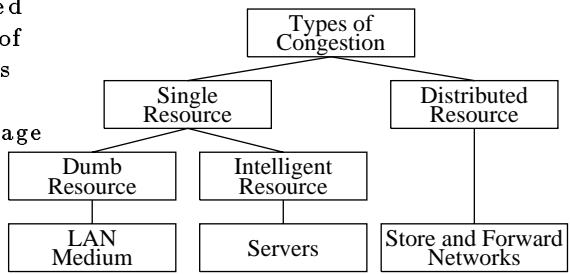

Figure 4: Types of congestion problems.

1. **Resource Creation Schemes**: Such schemes increase the capacity of the resource by dynamically reconfiguring them. Examples of such schemes are:

   - Dial-up links that can be added only during high usage.
   - Power increases on satellite links to increase their bandwidths.
   - Path splitting so that extra traffic is sent via routes that may not be considered optimal under low load.

   With all of the above schemes, users of the resource do not need to be informed, as they may not even be aware of the congestion in the network. The network is solely responsible for solving the congestion problem.

2. **Demand Reduction Schemes**: These schemes try to reduce the demand to the level of the available resources. Most of these schemes require that the user (or other control points) be informed about the load condition in the network so they can adjust the traffic. There are three basic classes of such schemes:

   - *Service Denial Schemes*: These schemes do not allow new sessions to start up during congestion. The busy tone provided by the telephone company is an example of such a scheme. Connection-oriented computer networks also use similar schemes where congestion at any intermediate node would prevent new sessions from starting up.
   - *Service Degradation Schemes*: These schemes ask all users (existing as well as new users) to reduce their loads. Dynamic window schemes in which the users increase or decrease the number of packets outstanding in the network based on the load are examples of this approach.

- *Scheduling Schemes*: These schemes ask users to schedule their demands so that the total demand is less than the capacity. Various contention schemes, and polling, priority, and reservation schemes are examples of this approach. It must be pointed out that all scheduling schemes are a special case of the service degradation approach.

In connectionless networks, starting a new session does not require that all intermediate resources be informed, so the service denial approach cannot be effectively used. Such networks generally use service degradation and scheduling techniques.

All congestion control schemes, resource creation as well as demand reduction schemes, require the network to measure the total load on the network and then to take some remedial action. The first part is often called **feedback**, while the second part is called **control**. Depending upon the load, a feedback signal is sent from the congested resource to one or more control points, which then take remedial action. In demand reduction schemes, the control point is generally the source node of the traffic, while in resource creation schemes, the control points may be other intermediate nodes (or sources) on the network. A number of feedback mechanisms have been proposed, for example:

- *Feedback Messages*: Explicit messages are sent from the congested resource to the control point. Such messages have been called choke packets, source quench messages, or permits. The sources reduce their loads upon the receipt of choke packets [24] or source quench messages and increase it if these are not received. In the isarithmic scheme [6], the sources have to wait to receive a permit before sending a packet. Critics of this approach argue that the extra traffic created by the feedback messages and permits during heavy load may worsen the congestion.

- *Feedback in Routing Messages*: Each intermediate resource sends its load level (typically in terms of queue length or delay) to all neighboring nodes who then adjust the level of traffic sent to that resource. The delay adaptive routing used in ARPAnet at one time is an example of this approach. This method was found to generate too many routing messages, since the rate of change of delay through a node was much faster than the rate at which control could be affected.

- *Rejecting Further Traffic*: In this approach, no explicit messages are sent. However, incoming packets are either lost or not acknowledged, thereby, creating a backpressure. This results in queues being built at other nodes, which then backpressure their neighbors. The backpressure slowly travels towards the source. This technique is useful only if the congestion lasts for a very short duration. Otherwise, the traffic that is not even using the congested resources is unfairly affected by the backpressure propagating throughout the network.

- *Probe Packets*: This requires sources to send probe packets through the network and to adjust their loads depending upon the delay experienced by the probe packets.

- *Feedback Fields in Packets*: This approach avoids the overhead caused by feedback messages by including the feedback in a special field in all packets. The feedback may be included either in packets going in the reverse direction (towards the source of congesting traffic) [9, 29] or in the forward direction (towards the destination), which then relays the information back to the source [19].

A number of alternatives for the location of control have also been proposed:

- *Transport Layer*: The traffic is generated by the end systems, therefore, they are in the best position to adjust the load in an efficient manner. Dynamic window schemes are an example of such controls at the transport layer. If the network and the end systems are under different administrative control, such as in public networks, the control may be exercised between the first and the last intermediate systems (entry-to-exit or DCE-to-DCE) instead of between the end systems.

- *Network Access*: Like traffic lights at the entrance ramps of some highways, the access controls at the network layer of the source node allow new traffic to enter the network only if the network is not congested. For example, the input limit scheme [23] does this by setting appropriate limits on buffers allocated to the traffic originating at the node and to the transit traffic.

- *Network Layer*: The routers and gateways, if congested, can take immediate action by reducing service to the sources that are sending more than their fair share. The fair queueing scheme [7], various buffer class schemes, and the leaky bucket algorithm [31] are examples of this approach. These schemes are particularly useful for public networks, which may not be able to ensure that the end systems will reduce the load on a congestion feedback signal.

- *Data Link Layer*: The control can also be exercised at the data link level at each hop using data link level flow control mechanisms. Backpressure on buffer exhaustion [3] is one such scheme.

There are a number of other policies at the transport, network, and data link layers that can be helpful in congestion control. These policies are discussed later in Section 5.

## 4 Why Is the Problem Difficult?

Despite the fact that a number of schemes have been proposed for congestion control, the search for new schemes continues. The research in this area has been going on for at least two decades [10]. There are two reasons for this. First, there are requirements for congestion control schemes that make it difficult to get a satisfactory solution. Second, there are several network policies that affect the design of a congestion scheme. Thus, a scheme developed for one network may not work on another network with a different architecture. In this section, we elaborate on the first issue of requirements. The second issue of network policies is discussed in the next section.

*The scheme must have a low overhead.* In particular, it should not increase traffic during congestion. This is one of the reasons why explicit feedback messages are considered undesirable. Some researchers have suggested that feedback be sent only during low load, thus, the absence of feedback would automatically indicate a high load. Even such schemes are not desirable, since the network resources are also used for nonnetworking applications. Therefore, resources consumed to process these additional messages could have been better used by these other applications.

*The scheme must be fair.* Fairness may not be important during low load when everyone's demands can be satisfied. However, during congestion when the resources are less than the demand, it is important that the available resources be allocated fairly. Defining fairness is not trivial. A number of definitions have been proposed [1, 11, 15, 16]. However, no one definition has been widely accepted. For example, some researchers consider starvation of a few users to be unfair [1]. Not allocating any resources to a user is called starvation. By this definition, if all users get a nonzero share of the resources, the scheme is fair. Others argue that a scheme without starvation can still be unfair if the resources are allocated unevenly. The key problem is in defining what is an even distribution of resources in a wide-area network where different users are traveling different distances. Some want to give preference to traffic that has traveled a long distance (more hops), while others want to give equal throughput to all users. The definition of users is also not clear. Some researchers treat each source-destination pair as a user. Giving equal throughput to all source-destination pairs passing through an intermediate node does not automatically guarantee that all connections from a single user will be treated fairly.

*The scheme must be responsive.* The available capacity on a network is a constantly changing quantity. As the nodes and links go up or down, the available capacity is increased or decreased. As the users start and stop, the demand also increases or decreases. The congestion control scheme is required to match the demand dynamically to the available capacity. Thus, it should ask users to increase the demand when additional capacity becomes available and to decrease it if the demand exceeds the capacity. The demand curve should follow the capacity curve very closely.

*The congestion scheme must work in bad environments.* Under congestion, the rate of transmission errors, out of sequence packets, deadlocks, and lost packets increases considerably. The congestion scheme must continue to work in spite of these conditions.

Finally, *the scheme must be socially optimal.* That is, the scheme must allow the total network performance to be maximized. Schemes that consider each user in isolation may be individually optimal, but not socially optimal [30, 21]. For example, if each user attempted to maximize its throughput, it may lead to an unstable situation where total network load keeps increasing.

It should be clear from the above list of requirements that designing a congestion control scheme is not a trivial problem

## 5 Policies That Affect the Congestion Control Scheme

Any architectural or implementation decision that affects either side of Equation 1 affects the design of a congestion control scheme. Thus, any design decision affecting the load (demand) or resource allocation can be considered a part of the overall congestion control strategy of the network. These decisions are called policies in this paper. A list of such policies is presented in Table I.

The most important network policy is the connection mechanism There are two types of networks: connection-oriented and connectionless. In connection-oriented networks, when a new session is set up, each

Table I. Policies That Affect Congestion

1. Network Layer:
   - Connection mechanism
   - Packet queuing and service policy
   - Packet drop policy
   - Packet routing policy
   - Lifetime control policy

2. Transport Layer:
   - Round-trip delay estimation algorithm
   - Timeout algorithm
   - Retransmission policy
   - Out-of-order packet caching policy
   - Acknowledgment policy
   - Flow control policy
   - Buffer management policy

3. Data Link Layer:
   - Data link level retransmission policy
   - Data link level queuing and service policy
   - Data link level packet drop policy
   - Data link level acknowledgment policy
   - Data link level flow control policy

intermediate node in the path to be used is asked to reserve certain resources for the session. If the resources are not available, the session is not started. In connectionless networks, new sessions can be started without any resource reservations at the intermediate nodes. This also allows the flexibility to dynamically change the paths of existing connections. It is clear that the service denial schemes will work in connection-oriented networks, but not in connectionless networks. Similarly, path splitting, if required, should be set up at session start up time in connection-oriented networks. While in connectionless networks, it can be dynamically started and stopped during a session.

Packet queuing and service policies in the intermediate nodes affect resource allocation among users. An intermediate node may have separate queues for each output link, each input link, or a combination of the two [23]. In some networks, there is a separate queue for each source and, thus, fairness among all sources can be guaranteed. However, this does not ensure fairness among users from the same source going to different destinations. If a separate queue is maintained for each source-destination pair, fairness among users from the same source to different destinations can be obtained. Several schemes to efficiently maintain and service such queues have been proposed. One scheme is to serve queues in a round-robin order [12]. This will cause the queues with large packets to get a larger share of the bandwidth than those with small packets. Schemes to tackle this inequity have also been proposed [7].

The packet drop policy deals with the issue of which packet is dropped if there is insufficient buffer space in a queue. Some of the alternatives are the first packet in the queue, the last packet in the queue (the arriving packet), or a randomly selected packet. The choice depends upon the type of application. For real-time communications, the older the message, the less valuable it is. Therefore, it is better to drop packets at head of the queue. This type of traffic has been called 'milk' and is contrasted with file and terminal traffic, which has been called 'wine' because older messages are more valuable than newer ones [5]. To ensure fairness, some have proposed random dropping, but others have argued its effectiveness [32].

The route selection policy, in general, and the path splitting policy, in particular, affect the resource allocation and, hence, congestion in the network. In most networks today, a low-speed path will be totally unused even if a parallel high-speed path is congested. Path splitting is performed only across paths of the same speed or across parallel links connecting the same nodes (one hop).

Lifetime control policies affect the length of time a packet stays in the network before being dropped. There may be too many unnecessary retransmissions (and, hence, load) if the lifetime is either too short or too long.

The round-trip delay estimation and the timeout interval computation algorithms used by the transport protocol also have a significant impact. In fact, finding a good algorithm for estimating round-trip delay in the presence of packet loss has been the first step towards finding a solution for congestion control [14, 17, 22]. Reducing the probability of false timeout alarms using the mean as well as the variance of the round-trip delay also improves the efficiency of congestion control mechanisms using timeouts [14].

The number of packets retransmitted on a packet loss affects the stability of timeout-based congestion schemes. The optimal number may depend upon the out-of-order packet caching policy at the destination. If the receiving transport does not cache out-of-order packets, loss of a single packet may require retransmission of the entire window. However, a comparison of several alternatives showed that if the packet loss is due to congestion, it is best to retransmit just one packet regardless of the caching policy at the destination.

The packet acknowledgment policy affects the feedback delay in congestion information reaching back to the source. If every packet is acknowledged, there may be too much traffic but the congestion feedback is fast. If some acknowledgments are withheld, the load due to acknowledgments is less, but the congestion feedback is delayed more.

The flow control policy used at the transport layer also affects the design of the congestion control scheme. For a comparison of various flow control policies see Maxemchuk and Zarki [25]. Briefly, there are two major classes of flow control schemes: window-based and rate-based. In a window-based scheme, the destination specifies the number of packets that a source can send. This helps solve the problem of buffer shortage at the destination. The source can further reduce the window in response to a congestion feedback signal from the network. In the rate-based scheme, the destination specifies a maximum rate in terms of packets per second or bits per second that the source is allowed to send. The current trend is towards rate-based flow control schemes.

The choice between window-based and rate-based flow control schemes depends partially upon the bottleneck resource at the destination. Memory capacity is measured by the number of packets that can be stored; the processing capacity is measured by the rate at which packets can be processed; link bandwidth is measured in terms of the number of bits per second that can be transmitted; and so on. Thus, if the destination is storing the received packets on a disk, it may be limited by the transfer rate of the disk, therefore, it is better to use a rate-based flow control schemes. On the other hand, if the destination has very little memory, it may want to use a window-based flow control scheme and limit the number of packets that it can receive at a time. Similar considerations apply in choosing the metric for expressing the rate. The choices are packets per second or bits per second. If the bottleneck or a similar device whose capacity is expressed in bits per second in the link, the rate limit should be specified in bits per second. On the other hand, if the bottleneck device is a processor which takes a fixed amount of time per packet regardless of the size, the rate should be expressed in packets per second.

Buffer management policy at the destination transport also affects the rate at which the packets can be accepted at the destination and, hence, the congestion level in the network [13]. The buffers may be located in the system space or user space. They may be shared or nonshared. Buffers may be one size or multiple sizes. The credit allocation policy may be pessimistic or optimistic. In the pessimistic case, the sum of all the windows permitted by that node will never be greater than the available space. In an optimistic scheme, the node will allocate more windows than available buffer space. This allows a higher throughput with a smaller probability of loosing the packets in cases where all the windows are being used. If the buffers are located in user space, sharing and optimism are less likely than if they are in system space.

The data link level policies are similar to the transport layer policies except that they apply to each hop in the network. For example, the intermediate systems in the network may have their own packet caching, acknowledgments, retransmission, and flow control policies. All of these will affect the design of the congestion control scheme.

In summary, there are a large number of architectural decisions that affect the design of a congestion control scheme. This is why analysts comparing the same set of alternatives may reach different conclusions. A scheme that works for one network may not work equally well for other networks. Some parameters or details of the scheme may have to be changed.

## 6 A Fundamental Principle of Control

As the name indicates, the problem of congestion control is basically a control problem. Most congestion control schemes consist of a feedback mechanism and a control mechanism. In control theory, it is well known that control frequency should be equal to the feedback frequency. As shown in Figure 5, if the control is faster than the feedback, the system will have oscillations and instability. On the other hand, if the control is slower than the feedback, the system will be tardy and slow to respond to changes. In designing congestion schemes it is important to apply this principle and to carefully set the control interval. In many existing schemes this is ignored, and although a feedback mechanism such as the source quench is specified, the issue of how often to send feedback and how long to wait before acting is left unspecified. This leads to schemes that are later found ineffective.

Another lesson to learn from the control theory principle is that no scheme can solve congestion that last less than its feedback delay. Transport level controls, such as dynamic window (or rate) schemes, work only if the con-gestion lasts for a few round-trip delays. For congestion that lasts for a shorter duration, data link and network level controls, such as priority classes, buffer classes, and input buffer limiting, are required. For longer-term congestion, either a session level control (such as session denial) or a resource creation scheme discussed earlier should be used. If congestion lasts indefinitely, it is best to solve the problem by installing extra resources. Dynamic schemes are good only for transient congestion. Also, since the duration of congestion can not be determined in advance, it is best to use a a combination of schemes operating at different layers.

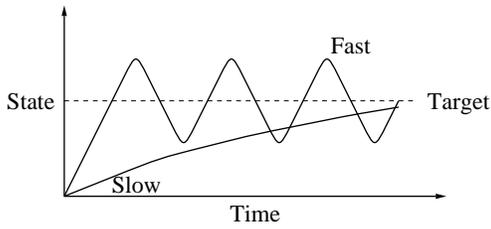

Figure 5: The rate of control and feedback delay are related.

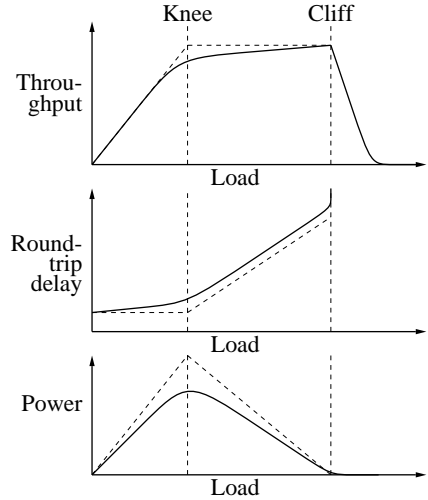

Figure 6: Network performance as a function of the load. Broken curves indicate performance with deterministic service and interarrival times.

## 7 Our Recent Proposals

In this section, we briefly describe three congestion schemes that we have recently proposed.

### 7.1 Timeout-Based Congestion Control

The timeout-based congestion control schemes are based on the idea that packet loss is a good indicator of congestion and, therefore, on a timeout, the load on the network should be reduced. Later, if there is no further loss, the load is increased slowly. In one timeout-based scheme called CUTE (Congestion Using Timeout at the End-to-end layer), the window is decreased to one on a timeout, and only one packet is retransmitted regardless of the window. Later, the window is increased from W to W+1 after receiving acknowledgments for W packets without any timeouts. The window versus the number of packets acknowledged in this case follows a parabolic curve and, therefore, this increase policy is called a parabolic increase. The complete scheme is described in Jain [18]. In a similar scheme by Bux [2], the window is increased linearly, that is by one after every eight packets. Recently, Jacobson [14] proposed another version called 'slow start' where the window $W$ at timeout is remembered, and the increase is linear up to $W_o/2$ and parabolic thereafter. Other combinations, such as decreasing to $W/2$ and increasing linearly after every five packets, have also been proposed [8].

### 7.2 DECbit Scheme for Congestion Avoidance

Another recent development in the area of congestion control is the introduction of the concept of **congestion avoidance**. Figure 6 shows general patterns of response time and throughput of a network as the network load increases. If the load is small, throughput generally keeps up with the load. As the load increases throughput increases. After the load reaches the network capacity, throughput stops increasing. This point is called the **knee**. If the load is increased any further, the queues start building, potentially resulting in packets being dropped. Throughput may suddenly drop when the load increases beyond this point. This point is called the **cliff** because the throughput falls off rapidly after this point.

A scheme that allows the network to operate at the knee is called a congestion avoidance scheme as distinguished from a congestion control scheme, which tries to keep the network operating in the zone to the left of the cliff. A simple congestion avoidance scheme using a single bit in the network layer header is summarized in [19] and described in further detail in [4, 20, 27, 28].

## 7.3 Delay-Based Scheme for Congestion Avoidance

One problem with schemes requiring explicit feedback from the network is that they cannot be used on heterogeneous networks that consist of networks with several different architectures. Since all the major networks of the world are slowly becoming interconnected, a packet may traverse several different types of networks before arriving at the destination. In such cases, the feedback provided by one network may not be meaningful to sources on other networks. Also, some intermediate nodes, for example, bridges, are susceptible to congestion, but cannot let their presence be known. In such cases, only schemes with implicit feedback can be used. The timeout-based scheme described earlier is an example of an implicit feedback scheme for congestion control. To achieve congestion avoidance using implicit feedback schemes is currently an unsolved problem. One tentative proposal calls for measuring delay and adjusting the traffic depending upon the delay [21]. More research in this area is required before this proposal can be implemented into networks.

All three schemes discussed in this section have two key features. First, they do not require any additional packets. As discussed earlier, processing of packets is expensive, and any attempt to increase network performance by introducing more packets may not be fruitful. Second, all parameters of the schemes are dimensionless. In particular, the schemes do not use any timers. The correct value for any timer depends upon the network size and the link speed. A scheme without any dimensional parameters is applicable to a wider range of link speeds and network sizes.

## 8 Areas for Further Research

Although congestion control is not a new problem, there are considerable opportunities for research. In this section, we point out several issues that need to be resolved.

*Path splitting among long paths of differing capacities* is not well understood. In most networks today, traffic from a given source to a given destination either passes through the same path or is split equally among different paths of equal capacities. Thus, if the optimal path is congested and a slower path is available, the slower path is not used. Designing a scheme that allows slower paths to be used depending upon the load levels on all paths is a topic for further research.

*Insulating one level of network hierarchy from congestion in other levels* is another area for research. Most large networks are organized hierarchically into several levels. Schemes are required that prevent congestion at one level from affecting the traffic at other levels. Thus, congestion of a backbone network should not affect other networks and vice versa.

*Congestion control in integrated networks* with voice, data, and several other types of traffic is also an interesting research problem. Giving higher priority to voice traffic, a commonly proposed solution, does not suite all environments. In some cases, such as real-time applications, the delay and throughput requirements are complex, and accommodating them in a congestion control scheme is nontrivial. As the telecommunication industry is moving towards asynchronous transfer mode (ATM), which uses short, fixed-size packets (cells), the congestion control schemes for such networks are being heatedly debated in several standards committees.

*Heterogeneous networks consisting of networks using several different architectures* need implicit feedback schemes for congestion control and avoidance. This problem was mentioned earlier.

*Dynamic link creation schemes that require the dialing up of a new link* need to be developed. When a link should be dialed up or disconnected depends upon the tariff structure. Now that high-speed, dial-up links are becoming available, it would be interesting to have guidelines regarding their usage.

*Server congestion* is a recent problem that started occurring with the introduction of distributed systems. After a power failure, all nodes in a building need access to the name server, boot server, and so on. Unless the access is regulated properly, the server can get congested with requests and may be so late in responding that the requests are retransmitted, thus causing an unnecessary additional load on the servers. Schemes to solve this problem need to be developed.

## 9 Summary

Congestion is not a static resource shortage problem; rather it is a dynamic resource allocation problem. Simply placing more memory in the nodes, or creating faster links or faster processors will not solve the congestion problem. In any intermediate system where the total input rate is higher than the output rate, queues will build up. Therefore, explicit measures to ensure that the input rate is reduced should be built into the protocol architectures.

Congestion occurs whenever the total demand is more than the total available resources of memory, links, processors, and so on. Therefore, congestion schemes can be classified as resource creation schemes or demand reduction schemes. Demand reduction schemes can be further subdivided into service denial, service degradation, and scheduling schemes. Several schemes that feedback the network load information to the sources, who in turn control traffic, have been proposed.

Congestion control is not a trivial problem because of the number of requirements, such as low overhead, fairness, responsiveness, and so on. In particular, congestion schemes are called to work under unfavorable network conditions and are required to ensure that the result is socially optimal.

A number of network policies affect the choice of congestion control schemes. This is why one scheme may not be suitable for all networks. Given a set of protocol design decisions, the congestion control scheme has to be tuned to work appropriately with that set.

One principle that is often ignored in quickly designed congestion control schemes is that the control and feedback rates should be similar. Otherwise, the system will have oscillatory or irresponsive behavior. This is why a combination of schemes working at data link, networking, and transport layers are required, along with proper capacity planning to overcome congestion lasting a short duration to a very long duration.

Finally, as the networks become larger and heterogeneous, with higher speeds and integrated traffic, the congestion problem becomes more difficult to handle and more important than ever.

## References

[1] K. Bharat-Kumar and J. M. Jaffe, "A New Approach to Performance-Oriented Flow Control," IEEE Transactions on Communications, Vol. COM-29, No. 4, April 1981, pp. 427-435.

[2] W. Bux and D. Grillo, "Flow Control in Local-Area Networks of Interconnected Token Rings," IEEE Transactions on Communications, Vol. COM-33, No. 10, October 1985, pp. 1058-66.

[3] D. Cheriton, "Sirpent: A High Performance Internetworking Approach," Proc. ACM SIGCOMM 89 Symposium on Communications Architectures and Protocols, Austin, TX, September 1989, pp. 158-169.

[4] D. M. Chiu and R. Jain, "Analysis of Increase and Decrease Algorithms for Congestion Avoidance in Computer Networks," Computer Networks and ISDN Systems, Vol. 17, 1989, pp. 1-14.

[5] D. Cohen, "Flow Control for Real-Time Communication," Computer Communication Review, Vol. 10, No. 1-2, January/April 1980, pp. 41-47.

[6] D. W. Davies, "The Control of Congestion in Packet-Switching Networks," IEEE Trans. Commun., Vol. COM-20, No. 6, June 1972.

[7] A. Demers, S. Keshav, and S. Shenker, "Analysis and Simulation of a Fair Queueing Algorithm," Proc. ACM SIGCOMM 89 Symposium on Communications Architectures and Protocols, Austin, TX, September 1989, pp. 1-12.

[8] B. T. Doshi and H. Q. Nguyen, "Congestion Control in ISDN Frame-Relay Networks," AT&T Technical Journal, November/December 1988, pp. 35-46.

[9] F. D. George and G. E. Young, "SNA Flow Control: Architecture and Implementation," IBM Systems Journal, Vol. 21, No. 2, 1982, pp. 179-210.

[10] M. Gerla and L. Kleinrock, "Flow Control: A Comparative Survey," IEEE Transactions on Communications, Vol. COM-28, No. 4, April 1980, pp. 553-574.

[11] M. Gerla, H. W. Chan, and J. R. Boisson de Marca, "Fairness in Computer Networks," Proc. IEEE International Conference on Communications ICC'85, Chicago, IL, June 23-26, 1985, pp. 43.5.1-6.

[12] E. L. Hahne and R. G. Gallager, "Round Robin Scheduling for Fair Flow Control in Data Communications Networks," Proc. IEEE International Conference on Communications ICC'86, Toronto, Canada, June 22-25, 1986, pp. 4.3.1-5.

[13] M. Irland, "Buffer Management in a Packet Switch," IEEE Trans. on Commun., Vol. COM-26, March 1978, pp. 328-337.

[14] V. Jacobson, "Congestion Avoidance and Control," Proc. ACM SIGCOMM 88, Stanford, CA, August 1988, pp. 314-329.

[15] J. M. Jaffe, "Bottleneck Flow Control," IEEE Transactions on Communications, Vol. COM-29, No. 7, July 1981, pp. 954-962.

[16] R. Jain, D. M. Chiu, and W. Hawe, *A Quantitative Measure of Fairness and Discrimination for Resource Allocation in Shared Systems*, Digital Equipment Corporation, Technical Report DEC-TR-301, Sept. 1984, 37 pp.

[17] R. Jain, "Divergence of Timeout Algorithms for Packet Retransmissions," *Proc. 5th Annual International Phoenix Conf on Computers and Communications*, Scottsdale, AZ, pp. 174-179, Mar. 1986.

[18] R. Jain, "A Timeout-Based Congestion Control Scheme for Window Flow-Controlled Networks," IEEE Journal on Selected Areas in Communications, Vol. SAC-4, No. 7, October 1986, pp. 1162-1167.

[19] R. Jain, K. K. Ramakrishnan, and D. M. Chiu, *Congestion Avoidance in Computer Networks with a Connectionless Network Layer*, Digital Equipment Corporation, Technical Report, DEC-TR-506, August 1987, 17 pp. Also in C. Partridge, Ed., *Innovations in Internetworking*, Artech House, Norwood, MA, 1988, pp. 140-156.

[20] R. Jain and K. K. Ramakrishnan, "Congestion Avoidance in Computer Networks with a Connectionless Network Layer: Concepts, Goals, and Methodology," Proc. IEEE Computer Networking Symposium, Washington, D. C., April 1988, pp. 134-143.

[21] R. Jain, "A Delay-Based Approach for Congestion Avoidance in Interconnected Heterogeneous Computer Networks," Computer Communications Review, Vol. 19, No. 5, October 1989, pp. 56-71.

[22] P. Karn and C. Partridge, "Improving Round-Trip Time Estimates in Reliable Transport Protocols," Proc. ACM SIGCOMM 87, Stowe, VT, August 1987, pp. 2-7.

[23] S. S. Lam and Y. C. Luke Lien, "Congestion Control of Packet Communication Networks by Input Buffer Limits – A Simulation Study," IEEE Transactions on Computers, Vol. C-30, No. 10, October 1981, pp. 733-742.

[24] J. C. Majithia, M. Irland, J. L. Grange, N. Cohen, and C. O'Donnell, "Experiments in Congestion Control Techniques," Proc. Int. Symp. Flow Control Computer Networks, Versailles, France. February 1979, pp. 211-234.

[25] N. F. Maxemchuk and M. E. Zarki, "Routing and Flow Control in High Speed, Wide Area Networks," Proceedings of IEEE, to appear.

[26] J. Nagle, "On Packet Switches with Infinite Storage," IEEE Transactions on Communications, Vol. COM-35, No. 4, April 1987, pp. 435-438.

[27] K. K. Ramakrishnan, D. M. Chiu, and R. Jain, *Congestion Avoidance in Computer Networks with a Connectionless Network Layer. Part IV: A Selective Binary Feedback Scheme for General Topologies*, Digital Equipment Corporation, Technical Report DEC-TR-510, August 1987, 41 pp.

[28] K. K. Ramakrishnan and R. Jain, "An Explicit Binary Feedback Scheme for Congestion Avoidance in Computer Networks with a Connectionless Network Layer," Proc. ACM SIGCOMM 88, Stanford, CA, August 1988, pp. 303-313.

[29] M. Schwartz, "Performance Analysis of the SNA Virtual Route Pacing Control," IEEE Transactions on Communications, Vol. COM-30, No. 1, January 1982, pp. 172-184.

[30] S. Stidham, Jr. "Optimal Control of Admission to a Queueing System," IEEE Transactions on Automatic Control, Vol. AC-30, No. 8, August 1985, pp. 705-713.

[31] J. S. Turner, "New Directions in Communications (or Which Way to the Information Age?)," IEEE Communications Magazine, Vol. 24, No. 10, October 1986, pp. 8-15.

[32] L. Zhang, "A New Architecture for Packet Switching Network Protocols," Ph.D. Thesis, Laboratory for Computer Science, Massachusetts Institute of Technology, Cambridge, MA, July 1989, 138 pp.